\documentclass[a4paper,12pt]{article}
\usepackage{amsfonts}
\usepackage{amssymb}
\usepackage{amsmath}
\usepackage{indentfirst}
\usepackage{dsfont}
\usepackage[latin1]{inputenc}
\def\be{\begin{equation}}
\def\ee{\end{equation}}
\def\bea{\begin{eqnarray}}
\def\eea{\end{eqnarray}}

\title{Twisted Fermionic Oscillator Algebra in $\kappa$-Minkowski space-time}
\author{Ravikant Verma\footnote{Email:ravikant.uohyd@gmail.com}~\\
{\it School of Physics, University of Hyderabad},\\{\it Central University P O, Hyderabad-500046, India}}

\begin{document}
\maketitle
\begin{abstract}
In this paper, we investigate the twisted algebra of the fermionic oscillators associated with Dirac field defined in $\kappa$-Minkowski space-time. Starting from $\kappa$-deformed Dirac theory, which is invariant under the undeformed $\kappa$-Poincare algebra, using the twisted flip operator, we derive the deformed algebra of the creation and annihilation operators corresponding to the Dirac field quanta in $\kappa$-Minkowski space-time. In the limit $a\rightarrow$0, the deformed algebra reduces to the commutative result.
\end{abstract}

\section{Introduction}

The noncommutative space-time was first introduced by Snyder in 1947\cite{sny} as suggested by Heisenberg, with the aim of handling the UV divergences in quantum field theories. In recent times, there were many developments in quantum gravity and string theory, where this notion of space-time whose coordinates do not commute appeared naturally at the microscopic level\cite{2a,2b,2c,2d} and in this approach, coordinates of the space-time get quantized. This leads to the modification of the symmetry of the space-time. One such deformed symmetry algebra that has attracted wide attention in the study of quantum gravity is the $\kappa$-Poincare algebra and the associated space-time is called the $\kappa$-Minkowski space-time.

Different aspects of $\kappa$-deformed space-time and associated symmetry algebra have been studied extensively in recent times\cite{c1,c2,c3,c4,c5,c6}. In\cite{11,12}, $\kappa$-deformed Klein-Gordon equation was constructed and analysed. It was shown that the quantized field operators satisfy a deformed oscillator algebra. By applying this $\kappa$-deformed Klein-Gordon field to a uniformly accelerating detector, the modification of Unruh effect in $\kappa$-space-time was studied in\cite{eh1}. This analysis was extended to the case of $\kappa$-deformed Dirac field in\cite{eh}. The $\kappa$-Dirac equation is constructed as in commutative space-time, by demanding that its square should be the $\kappa$-deformed Klein-Gordon equation. But the Dirac equation obtained in\cite{27,28,29,30} was not invariant under the spin-half representation of $\kappa$-Poincare algebra, whereas this problem was solved in\cite{54}. But the square of the $\kappa$-Dirac equation obtained in \cite{54} was related to the $\kappa$-
deformed Pauli-Lubanski vector in contrast to the situation in commutative space-time. Using other approaches the Dirac equation in $\kappa$-space-time was constructed and studied in\cite{55,56,57,58,59,60,61}. In this paper, we use $\kappa$-Dirac equation obtained in\cite{nhpl}, which is invariant under the undeformed $\kappa$-Poincare algebra. This $\kappa$-Dirac equation is constructed by using the Dirac derivative which transform like a 4-vector under the undeformed $\kappa$-Poincare algebra. The generators of the undeformed $\kappa$-Poincare algebra defined in terms of the commutative coordinates and corresponding derivatives\cite{fgrd1,fgrd2,fgrd3,fgrd4}. 

The statistics and spin of quantum fields and corresponding particles do have important implications and hence of intrinsic interest. Thus, it is natural to ask how these notions are affected by noncommutativity of the space-time. This has been studied for the various types of noncommutative space-times in\cite{5,6,7,8,9,10}. The noncommutative space-time do modify the symmetry algebra of the underlying space-time and it is known that the corresponding symmetry is described by a Hopf algebra. In the Hopf algebra approach, the symmetry algebra of the noncommutative space-time acts on the multiparticle states through the twisted coproduct of the generators of the algebra. This twist of the coproduct is necessitated by the requirement that the action of the symmetry algebra is compatible with the noncommutativity of the coordinates of the underlying space-time. This modification or deformation of the coproduct do affect the notion of statistics\cite{6}. Conventionally, the statistics of multiparticle system is 
defined by the transformation property of the states under flipping of two particles. This action of flip operator and the action of underlying symmetry algebra should be compatible and this requirement is guaranteed since the coproduct and the flip operator commute in case of theories in the commutative space-time. Since in the noncommutative space-time, the symmetry is realized through the Hofp algebra whose coproduct is twisted, the usual flip operator do not commute with it\cite{6,7,8,9}. This leads to the modification of the statistics of the field quanta. For the particles described by $\kappa$-deformed Klein-Gordon equation this deformed statistics was obtained in\cite{11}. Here, by demanding that the flip operator should commute with the twisted coproduct, a deformed coproduct was obtained and a twisted flip operator that commute with this twisted coproduct was also derived. Then by demanding that the multiparticle states of the $\kappa$-deformed scalar particle should be symmetric under the twisted 
flip operator, it was shown that the creation and annihilation operators corresponding to the $\kappa$-deformed Klein-Gordon theory satisfy the deformed oscillator algebra. In this paper, we extend this study to the $\kappa$-deformed Dirac quanta and derive the deformed algebra of the creation and annihilation operators corresponding to the field quanta described by $\kappa$-Dirac fermions.

This paper is organized as follows. In the next section, we give a brief summary of $\kappa$-Minkowski space-time and in section 3, we summarise the $\kappa$-Dirac equation obtained in\cite{nhpl}, required for our analysis. In section 4, we recall twisted flip operator for $\kappa$-Minkowski space-time obtained in\cite{11}. Our main results are presented in section 5. Here, we derive the deformed algebra of creation and annihilation operators of $\kappa$-deformed Dirac field. We present our concluding remarks in section 6.

\section{$\kappa$-Minkowski space-time}

In this section, we present a brief summary of the $\kappa$-deformed space-time \cite{fgrd1,fgrd2,fgrd3,fgrd4}, which is an example of a noncommutative space-time whose coordinates obey Lie algebra type commutation relations, i.e.,
\begin{equation} 
[\hat{x}_\mu, \hat{x}_\nu]=i C_{{\mu}{\nu}{\lambda}}\hat{x}^\lambda.
\end{equation}
Here $C_{{\mu}{\nu}{\lambda}}=a_{\mu} \eta_{{\nu}{\lambda}} - a_{\nu} \eta_{{\mu}{\lambda}}$,~ $\eta_{{\mu}{\nu}}$ = diag(-1,1,1,1). Here $a_\mu(\mu=0,1,2,3)$ are real, dimensionfull constants and characterize the deformation of the Minkowski space-time. For the $\kappa$-Minkowski space-time, $a_i =0$, $i=1,2,3$ and $a_0=a=\frac{1}{\kappa}$. Thus we find the commutation relations between the coordinate of $\kappa$-Minkowski space-time as
 \begin{equation}
 [\hat{x}^i, \hat{x}^j]=0,~~ [\hat{x}^0, \hat{x}^i]=i a \hat{x}^i,~~  a=\frac{1}{\kappa}.
 \end{equation}\\
Note that $a$ has the dimension of length. Using the Minkowski metric $\eta_{{\mu}{\nu}}$ = diag(-1,1,1,1) we define $x^\mu =\eta^{{\mu}{\alpha}} x_\alpha$ and $\partial^{\mu}=\eta^{{\mu}{\alpha}} \partial_\alpha$ which obeys the relations
\begin{eqnarray}
&[x_\mu, x_\nu]=0 ,~~ [\partial_\mu, \partial_\nu]=0 ,~~ [\partial^{\mu} , x_{\nu}]=\eta^{\mu} _{\nu}&.\\
&[\partial_\mu, x_\nu]=\eta_{{\mu}{\nu}}, [p_\mu , x_\nu]=- i \eta_{{\mu}{\nu}},p_\mu=- i \partial_\mu &.
\end{eqnarray}
The $\kappa$-Minkowski space-time coordinates can be expressed in terms the coordinates of commutative space-time $x_\mu$ and corresponding derivatives $\partial_\mu$ as a power series as
\begin{equation}
\hat{x}_\mu=x^\alpha \phi_{{\alpha}{\mu}}(\partial).
\end{equation}
It is easy to see that these coordinate obeys $[\partial_\mu , \hat{x}_\nu]=\phi_{{\mu}{\nu}}(\partial)$. Explicitly, $\hat{x}_i$ and $\hat{x}_0$ are given as
\begin{eqnarray}
\hat{x}_i=x_i \varphi(A),~~ \hat{x}_0=x_0 \psi(A) + i a x_i \partial_i \gamma(A).\label{partial}
\end{eqnarray}
Where $A=- ia \partial_0$. Using equation (6) in equation (2), we find
\begin{center}
$\frac{\varphi^\prime}{\varphi}\psi=\gamma(A)-1,$
\end{center}
where $\varphi^\prime =\frac{d\varphi}{dA}$ satisfying boundary condition $\varphi(0)=1,~\psi(0)=1$ and $\gamma(0)=\varphi^\prime(0) + 1$ are finite and positive functions. Further demanding that commutator between generators $M_{\mu\nu}$ of undeformed $\kappa$-Poincare algebra and coordinates of $\kappa$-space-time must be linear in $\hat{x}_\mu$ and generators $M_{\mu\nu}$, one get
\begin{equation}
[M_{\mu\nu},\hat{x}_{\lambda}]=\hat{x}_{\mu}\eta_{\nu\lambda}-\hat{x}_{\nu}\eta_{\mu\lambda}-i(a_\mu M_{\nu\lambda}-a_\nu M_{\mu\lambda}).
\end{equation}
 We also demand that these commutators should have correct commutative limit. This leads to two type of possible realization, one where $\psi=1$ and second one where $\psi=1+2A$\cite{fgrd1}. Here we restrict ourself to the realization $\psi=1$. Thus, we find 
\begin{equation}
[\partial_i , \hat{x}_j]=\delta_{{i}{j}} \varphi(A) ,~~[\partial_i , \hat{x}_0]=i a \partial_{i} \gamma(A) ,~~ [\partial_0 , \hat{x}_i]=0,
\end{equation}
and
\begin{equation}
[\partial_0 ,\hat{x}_0]=\eta_{00}=-1.
\end{equation} 
For the realization $\psi=1$, the explicit form of $M_{\mu\nu}$ are
\begin{equation}
M_{ij}=x_{i}\partial_{j}-x_{j}\partial_{i}\label{dg1}
\end{equation}
\begin{equation}
M_{0i}=x_i \partial_0 \varphi \frac{e^{2A}-1}{2A}-x_0 \partial_i \frac{1}{\varphi}+iax_i \bigtriangledown^2 \frac{1}{2\varphi}-iax_k \partial_k \partial_i \frac{\gamma}{\varphi}.\label{dg2}
\end{equation}
Where  $\bigtriangledown^2=\partial_k \partial_k$. Here we note that in Minkowski space-time, generators of Poincare algebra are $\partial_0,\partial_i$ and $\tilde{M}_{\mu\nu}=x_{\mu}\partial_{\nu}-x_{\nu}\partial_{\mu}$. But it is clear that $\partial_\mu$ do not transform like a 4-vector under the transformations generated by $M_{\mu\nu}$ defined in Eqn.(\ref{dg1}) and Eqn.(\ref{dg2}). Therefore one introduces Dirac derivative $D_\mu$, which transform like a 4-vector under the algebra generated by $M_{\mu \nu}$ given in Eqn.(\ref{dg1}) and Eqn.(\ref{dg2}).
 
The symmetry algebra of the underlying $\kappa$-space-time  is known as the undeformed $\kappa$-Poincare algebra. Their generators $D_{\mu}$ and $M_{{\mu}{\nu}}$ obey\cite{fgrd1,fgrd2,fgrd3,fgrd4}
\begin{equation}
[M_{{\mu}{\nu}} , D_{\lambda}]=\eta_{{\nu}{\lambda}} D_\mu - \eta_{{\mu}{\lambda}} D_\nu ,~~ [D_\mu , D_\nu]=0,\label{dirac4}
\end{equation}
\begin{equation}
[M_{{\mu}{\nu}}, M_{{\lambda}{\rho}}]=\eta_{{\mu}{\rho}} M_{{\nu}{\lambda}} + \eta_{{\nu}{\lambda}} M_{{\mu}{\rho}} - \eta_{{\nu}{\rho}} M_{{\mu}{\lambda}} - \eta_{{\mu}{\lambda}} M_{{\nu}{\rho}}.\label{dirac5}
\end{equation}
We also note that the twisted coproduct of the rotation and boost generators\cite{fgrd1,fgrd2,4} are
\begin{equation}
\triangle_\varphi(M_{ij})=M_{ij}\otimes I+I\otimes M_{ij},
\end{equation}
\begin{equation}
\triangle_\varphi(M_{i0})=M_{i0}\otimes I+e^A\otimes M_{i0}+ ia\partial_i \frac{1}{\varphi(A)}\otimes M_{ij}.
\end{equation}
For realization $\psi=1$, explicit form of Dirac derivative $D_\mu$ are
\begin{equation}
D_i=\partial_i \frac{e^-A}{\varphi},~~ D_0=\partial_0 \frac{sinhA}{A} + i a \bigtriangledown^2 \frac{e^-A}{2 \varphi^2}. \label{dr}
\end{equation}
The algebra defined in Eqns.(\ref{dirac4},\ref{dirac5}) is same as of Poincare algebra. But it can be seen from Eqn.(\ref{dg1}), Eqn.(\ref{dg2}) and Eqn.(\ref{dr}) that the explicit form of generators are deformed compared to that of the generators of the Poincare algebra. Hence, this algebra defined by Eqn.(\ref{dirac4}) and Eqn.(\ref{dirac5}) is referred as the undeformed $\kappa$-Poincare algebra.

The Casimir of this undeformed $\kappa$-Poincare algebra is $D_{\mu}D^{\mu}$ and it can be expressed as 
\begin{equation}
D_{\mu}D^{\mu}=\square (1 - \frac{a^2}{4} \square).\label{casmir}
\end{equation}
The $\square$ operator in the above satisfy
\begin{equation}
[M_{{\mu}{\nu}} , \square]=0, [\square,\hat{x}_\mu]=2D_\mu ,
\end{equation}
and the explicit form of the $\square$ operator is
\begin{equation}
\square=\bigtriangledown^2 \frac{e^{-A}}{2 \varphi^2} + 2 \partial_0^2 \frac{(1-coshA)}{A^2},\label{box}
\end{equation}
where $\bigtriangledown^2=\partial_i \partial_i$ and $A=-i a \partial_0$. Note that $\partial_i$ and $\partial_0$ are the derivatives corresponding to the commutative space-time coordinates. It is clear that the Casimir, $D_{\mu}D^{\mu}$ reduces to the usual relativistic dispersion relation in limit $a\rightarrow0$. $\varphi$ appearing in above equations, characterizes arbitrary realization of the $\kappa$-space-time coordinates in terms of commutative coordinates and their derivatives.

\section{$\kappa$-deformed Dirac equation and it's solution}

The massless Klein-Gordon equation in the commutative space-time can be expressed as $p_\mu p^\mu \Phi(x)=0$, where $p_\mu p^\mu$ is the Casmir operator of the Poincare algebra. Generalizing this procedure to $\kappa$-space-time one obtains the $\kappa$-deformed Klein-Gordon equation\cite{11,12,fgrd1,fgrd2,fgrd3,fgrd4}. Thus the generalized Klein-Gordon equation, invariant under the $\kappa$-Poincare algebra, is written, using the Casimir of the undeformed $\kappa$-Poincare algebra as
\begin{equation}
\square (1 - \frac{a^2}{4} \square) \Phi(x) - m^2 \Phi(x)=0. \label{kg}
\end{equation}
It is clear from above that the scalar field and the operator appearing in the $\kappa$-deformed Klein-Gordon equation are defined in the commutative space-time itself.
\begin{sloppypar}
The deformed dispersion relation resulting from Eqn.(\ref{kg}) is
\end{sloppypar}
\begin{equation}
\frac{4}{a^2} sinh^2\left(\frac{ap_0}{2}\right) - p_{i}^{2} \frac{e^{-ap_0}}{\varphi^2(ap_0)} - \frac{a^2}{4} \left[\frac{4}{a^2} sinh^2\left(\frac{ap_0}{2}\right) - p_{i}^{2} \frac{e^{-ap_0}}{\varphi^2(ap_0)}\right]^2 = m^2.\label{D}
\end{equation}
Where $p_0=-i \partial_0$ and $p_i=- i \partial_i$. Now as in the commutative space-time, one constructs the Dirac equation by demanding that the square of Dirac equation should be the above Klein-Gordon equation. Thus one finds the $\kappa$-deformed Dirac equation as\cite{nhpl}
\begin{equation}
(i\gamma^0 D_0 + i\gamma^i D_i + m)\Psi=0 ,\label{24}
\end{equation}
where $D_0$, $D_i$ are the Dirac derivative defined in Eqn.(\ref{dr}). It is easy to see that the square of above equation gives the $\kappa$-deformed Klein-Gordon equation given in Eqn.(\ref{kg}), as required.

This $\kappa$-Dirac equation is invariant under the action of parity operator as well as time reversal operator\cite{nhpl}. The $\kappa$-Dirac equation for charged particle interacting with external electromagnetic field is obtained by replacing $p_\mu$ by $p_\mu - eA_\mu$, where $e$ is the electric charge of the particle and it was shown that this equation is not invariant under the charge conjugation\cite{nhpl}.

For the choice $\varphi=e^{-\frac{A}{2}}$, the field appearing in the $\kappa$-deformed Dirac equation (\ref{24}) can be expressed as
\be 
\Psi(x)=\int \frac{d^3p}{\sqrt{(2\pi)^3 2\omega}}\sum_{s=1,2}\left(b_s(p)u_s(p)e^{-ipx}+d_s^\dagger(p)v_s(p)e^{ipx}\right),\label{50}
\ee
\be 
\bar{\Psi}(y)=\int \frac{d^3q}{\sqrt{(2\pi)^3 2\omega^\prime}}\sum_{s\prime=1,2}\left(b_{s^\prime}^\dagger(q)\bar{u}_{s^\prime}(q)e^{-iqy}+d_{s^\prime}(q)\bar{v}_{s^\prime}(q)e^{iqy}\right).\label{51}
\ee
In the above two equations, the effect of $\kappa$-deformation is contained in $\omega$ and $\omega^\prime$ (see below for details). Here $b$, $d$, $b^\dagger$, $d^\dagger$ are the fermionic annihilation and creation operators, respectively. As shown in \cite{ravi}, the spinor representation under which the above $\kappa$-deformed spinors transform is characterised by $\sigma_{\mu\nu}=[\gamma_\mu,~\gamma_\nu]_+$ which is unaffected by $\kappa$-deformation. All the effects of the deformation enter the discussion of Lorentz transformation solely through the boost parameter. Other than this transformation parameter, the information about the $\kappa$-deformation is contained only in the explicit expression of the energy $\omega$ and $\omega^\prime$ appearing in Eqns.(\ref{50},~\ref{51}).

Now we show that the Dirac field given in Eqn.(\ref{50}) and Eqn.(\ref{51}) are the solution of Dirac equation which is given in Eqn.(\ref{24}).The $\kappa$-deformed Dirac equation given in Eqn.(\ref{24}) has the same form as the Dirac equation in the commutative space-time. The only difference is that the derivative operators appearing in the $\kappa$-deformed equation are the deformed operators given in Eqn.(\ref{dr}). Hence, it is easy to see that the plane wave solution to the $\kappa$-deformed Dirac equation will be same as in the commutative space-time. Thus the positive energy solution is of the form
\be 
\Psi(x) \backsim u_s (p) e^{-ip\cdot x},\label{a2}
\ee
and negative energy solution is
\be 
\Psi(x) \backsim v_s (p) e^{ip\cdot x}.\label{a3}
\ee
In above $u_s (p)$ and $v_s (p)$ are the positive and negative energy spinors. Now using Eqn.(\ref{a2}) and Eqn.(\ref{a3}) in Eqn.(\ref{24}) we get
\be 
(\gamma^\mu P_\mu + m)u_s (p) =0,\label{a4}
\ee
and
\be 
(-\gamma^\mu P_\mu + m)v_s (p) =0.
\ee
In the above $P_\mu =-iD_\mu$ and the explicit form of $D_0$ and $D_i$ given in Eqn.(\ref{dr}). It is easy to see that the Fourier transform of free field 
\be 
\Psi(x)=\frac{1}{(2\pi)^{3}}\int d^4p ~\delta(P^2 - m^2)\Theta(p_0) \left( b_s(p)u_s (p) e^{-ip\cdot x} + d_s^\dagger(p) v_s (p) e^{ip\cdot x}\right),
\ee 
where $P^2=P_\mu P^\mu$ satisfy the $\kappa$-deformed Dirac equation.

It is easy to see that the integration measure $\frac{d^3 p}{(2\pi)^3 2\omega}$ appearing in Eqn.(\ref{50}) is invariant under the undeformed $\kappa$-Poincare algebra. The dispersion relation in Eqn.(\ref{D}) is square of the $\kappa$-Dirac Eqn.(\ref{24}). For realisation $\varphi(A)=e^{-\frac{A}{2}}$, Eqn.(\ref{D}) becomes
\begin{equation}
{\cal A}^2 - \frac{4}{a^2} {\cal A} +\frac{4}{a^2} m^2 =0,
\end{equation}
where ${\cal A} =\frac{4}{a^2} sinh^2\left(\frac{ap_0}{2}\right) - p_{i}^{2} $.  From the above equation, we get
\begin{equation}
{\cal A} =\frac{2}{a^2} \pm \frac{2}{a^2} \sqrt{1-a^2 m^2}=H_\pm,
\end{equation}
Note that only $H_{-}$ has the correct commutative limit and hence we use only $H_{-}$ for the remaining calculations. i.e.,
\begin{equation}
\frac{4}{a^2} sinh^2\left(\frac{ap_0}{2}\right) - p_{i}^{2}=H_{-}.
\end{equation}
From above equation, we find
\begin{equation}
p_0=\omega =\pm \frac{2}{a} sinh^{-1}\left(\frac{a}{2}\sqrt{p_i^2 + H_{-}}\right).
\end{equation}
Note that we get commutative limit $p_0=\sqrt{p_i^2 + m^2}$, when $a\rightarrow 0$. Thus all the effect of $\kappa$-deformation is to modify the energy $p_0$ of the system, as expected. To see that $\int \frac{d^3 p}{(2\pi)^3 2\omega}$ is the invariant measure under the undeformed $\kappa$-Poincare algebra, we start form the four-dimensional momentum space (as in the commutative space-time) and carry out the integration over $p_0$ and obtain (with $p_0 >0$, positive energy condition),
\begin{eqnarray}
\int\frac{d^4 p}{(2\pi)^3}\delta(P^2 - m^2)\Theta(p_0)&=&\int\frac{d^4 p}{(2\pi)^3}\delta(p_0^2 - \omega^2)\Theta(p_0)\nonumber\\
&=&\int\frac{d^4 p}{(2\pi)^3}\delta[(p_0 - \omega)(p_0 + \omega)]\Theta(p_0)\nonumber\\
&=&\int\frac{d^4 p}{(2\pi)^3}\frac{1}{2|p_0|}\delta[(p_0 - \omega)+(p_0 + \omega)]\Theta(p_0)\nonumber\\
&=&\int\frac{d^3p dp_0}{(2\pi)^3 2|p_0|}\delta(p_0 - \omega)=\int\frac{d^3p}{(2\pi)^3 2\omega}.
\end{eqnarray}
This shows that the integrand of Eqn.(\ref{50}) is invariant under the undeformed $\kappa$-Poincare algebra.

\section{Twisted flip operator}

In this section, we present a brief summary of the twisted flip operator in the $\kappa$-deformed space-time\cite{11,12}. The coproduct $\bigtriangleup_\varphi$ of the partial derivatives in the realization given in Eqn.(\ref{partial}) are
\be
\triangle_\varphi(\partial_0)=\partial_0 \otimes I+I\otimes \partial_0\equiv \partial_{0}^{x}+\partial_{0}^{y}, \label{copro1}
\ee
\bea
\triangle_\varphi(\partial_i)&=&\varphi(A \otimes I+I\otimes A)\left[ \frac{\partial_i}{\varphi(A)}\otimes I+e^{A}\otimes \frac{\partial_i}{\varphi(A)} \right],\nonumber\\
&=&\partial_{i}^{x}\frac{\varphi(A_x +A_y)}{\varphi(A_x)}+\partial_{i}^{y}\frac{\varphi(A_x +A_y)}{\varphi(A_y)}e^{A_x}, \label{copro}
\eea
where $A_x=-ia\partial_{0}^{x}$ and $A_y=-ia\partial_{0}^{y}$. The twisted coproducts can be obtained from the untwisted ones by applying the twist element $\cal{F_\varphi}$. This twist element is the same as the one appearing in the definition of the star product. For any realization $\varphi$, the star product can be defined in terms of the twist element $\cal{F_\varphi}$ as
\be 
f\star_\varphi g=m_0( {\cal{F_\varphi}} f\otimes g)=m_\varphi (f\otimes g)\label{fg},
\ee
where $f$ and $g$ are the functions of the commutative coordinates and $m_0$ is the usual point wise multiplication map in the commutative algebra of smooth functions (and $m_0(f\otimes g)=fg$). The star product of two functions $f$ and $g$ in the $\varphi$ realization is given by
\be 
(f\star_\varphi g)(x)=m_0\left(e^{x_i(\bigtriangleup_\varphi - \bigtriangleup_0)\partial_i}f(u)\otimes g(t)\right)|_{u=t=x_i},
\ee
where $\bigtriangleup_\varphi$ is the twisted coproduct given in Eqn.(\ref{copro}) and the untwisted coproduct given by $\triangle_0(\partial)=\partial \otimes I+I\otimes \partial$. The corresponding twist element given by
\be 
{\cal{F_\varphi}}=e^{x_i(\bigtriangleup_\varphi - \bigtriangleup_0)\partial_i},\label{30}
\ee
where $\bigtriangleup_\varphi$ satisfies the relation
\be 
\bigtriangleup_\varphi = {\cal{F}}_{\varphi}^{-1} {\bigtriangleup_0} {\cal{F_\varphi}}.
\ee
Now using Eqns.(\ref{copro1}) and (\ref{copro}) in Eqn.(\ref{30}), we find
\be
{\cal{F_\varphi}}=e^{x_i\partial_{i}^{x}(\frac{\varphi(A_x +A_y)}{\varphi(A_x)}-1)+x_i\partial_{i}^{y}(\frac{\varphi(A_x +A_y)}{\varphi(A_y)}e^{A_x}-1)}.\label{32}
\ee
It is easy to show that\cite{11,12,4}
\be
\lim_{u\rightarrow x}e^{x_i\partial_{i}^{u}(\Phi(A_u , A_y)-1)} =e^{x_i\partial_{i}^{x}\ln \Phi(A_x , A_y)}.\label{id} 
\ee
Using above equation in Eqn.(\ref{32}), we find
\be
{\cal{F_\varphi}}=e^{N_x\ln \frac{\varphi(A_x +A_y)}{\varphi(A_x)}+N_y\left(A_x +\ln \frac{\varphi(A_x +A_y)}{\varphi(A_y)}\right)}
\ee
which can be written as
\be
{\cal{F_\varphi}}=e^{(N\otimes I)\ln \frac{\varphi(A\otimes I +I\otimes A)}{\varphi(A\otimes I)}+(I\otimes N)\left(A\otimes I +\ln \frac{\varphi(A\otimes I +I\otimes A)}{\varphi(I\otimes A)}\right)}\label{35}
\ee
where $N_x=x_i\partial_{i}^{x}$ and $N_y=x_i\partial_{i}^{y}$.

By using the explicit expression for the twist element $\cal{F_\varphi}$, one can see that the deformed flip operator is of the form
\be
\tau_\varphi= {\cal{F}}_{\varphi}^{-1} {\tau_0} {\cal{F_\varphi}}.\label{flip1} 
\ee
This deformed flip operator is independent of the function $\varphi$ that characterize the realization. Here $\tau_0$ is the undeformed flip operator associated with  exchange as
\be 
\tau_0 (f\otimes g)=+ g\otimes f,
\ee
where, $f$ and $g$ represent bosonic fields. If $f$ and $g$ represent fermionic fields, we will have 
\be 
\tau_0 (f\otimes g)=- g\otimes f.
\ee
The undeformed flip operator $\tau_0$ commute with the corresponding coproduct of symmetry group, i.e.,
\be
[\triangle_0 , \tau_0]=0, 
\ee
This shows that the statistics is unaffected by the action of the symmetry group.

In the $\kappa$-Minkowski space-time also it is desirable that the coproduct and flip operator commute. This requirement will provide a (anti)symmetrization procedure compatible with the action of twisted symmetry algebra in the $\kappa$-Minkowski space-time. It is easy to see that the flip operator $\tau_0$ does not commute with the twisted coproduct, i.e.,
\be
[\triangle_\varphi , \tau_0]\neq 0. 
\ee
From above it is clear that the symmetrization or antisymmetrization carried out by $\tau_0$, is not preserved by the action of the twisted symmetry group. Therefore, we must have a new flip operator, called the twisted flip operator $\tau_\varphi$ which satisfies the condition
\be
[\triangle_\varphi , \tau_\varphi]= 0. 
\ee
Straight forward calculation shows that the twisted flip operator in Eqn.(\ref{flip1}) satisfy the above requirement. Now we restrict to special case of $\varphi$ realization given in Eqn.(\ref{partial}). For this realization, we obtain the expression for twisted flip operator using Eqn.(\ref{35}) as
\be 
\tau_\varphi=e^{-A_x N_y +A_y N_x}\tau_0 \equiv e^{-A\otimes N +N\otimes A}\tau_0.
\ee
It is clear that in the limit $a\rightarrow 0$, we get back the correct commutative limit.

\section{Twisted Dirac Oscillator algebra}

In this section, we present the main result of this paper. Here, we obtain the twisted anti-commutators between the fermionic creation and annihilation operators that appear in the second quantized Dirac field in the $\kappa$-space-time.

Using the the twisted flip operator $\tau_\varphi$, now we define anti-symmetric states of a fermionic theory defined in $\kappa$-Minkowski space-time. We start by defining the deformed product of fermionic states as
\be 
f\star_\varphi g=m_\varphi(f\otimes g)=m_\varphi \tau_\varphi(f\otimes g).
\ee 
Here using the definition of $m_\varphi$ and $\tau_\varphi$ (see Eqn.(\ref{fg}) and Eqn.(\ref{flip1})), we get
\be
f\otimes g=\tau_\varphi (f\otimes g).\label{g}
\ee
Note that in the limit $a\rightarrow 0$, we get $f \otimes g = -g \otimes f$, as expected for fermionic fields $f$ and $g$. Now using the above equation for product of two fermionic fields, we find
\be 
\Psi(x)\otimes \Psi(y)+e^{-A\otimes N +N\otimes A}\Psi(y)\otimes \Psi(x)=0,\label{44}
\ee
\be 
\Psi(x)\otimes \bar{\Psi}(y)+e^{-A\otimes N +N\otimes A}\bar{\Psi}(y)\otimes \Psi(x)=\delta^3(x-y),
\ee
\be 
\bar{\Psi}(x)\otimes \bar{\Psi}(y)+e^{-A\otimes N +N\otimes A}\bar{\Psi}(y)\otimes \bar{\Psi}(x)=0.\label{49}
\ee
Now using Eqn.(\ref{50}) and Eqn.(\ref{51}) in Eqns.(\ref{44}-\ref{49}), we obtain various twisted anticommutation relations between creation and annihilation operators as
\be
b_s(p_0,\vec{p})b_{s^\prime}^\dagger(q_0,\vec{q})+e^{aq_0\frac{\partial}{\partial p_i}p_i+ap_0\frac{\partial}{\partial q_i}q_i}b_{s^\prime}^\dagger(q_0,\vec{q})b_s(p_0,\vec{p})=\delta^3(p-q)\delta_{ss^\prime},\label{52}
\ee
\be
d_s^\dagger(p_0,\vec{p})d_{s^\prime}(q_0,\vec{q})+e^{-aq_0\frac{\partial}{\partial p_i}p_i-ap_0\frac{\partial}{\partial q_i}q_i}d_{s^\prime}(q_0,\vec{q})d_s^\dagger(p_0,\vec{p})=\delta^3(p-q)\delta_{ss^\prime},
\ee
\be
b_s(p_0,\vec{p})b_{s^\prime}(q_0,\vec{q})+e^{-aq_0\frac{\partial}{\partial p_i}p_i+ap_0\frac{\partial}{\partial q_i}q_i}b_{s^\prime}(q_0,\vec{q})b_s(p_0,\vec{p})=0,
\ee
\be
d_s^\dagger(p_0,\vec{p})d_{s^\prime}^\dagger(q_0,\vec{q})+e^{aq_0\frac{\partial}{\partial p_i}p_i-ap_0\frac{\partial}{\partial q_i}q_i}d_{s^\prime}^\dagger(q_0,\vec{q})d_s^\dagger(p_0,\vec{p})=0,
\ee
\be
b_s^\dagger(p_0,\vec{p})b_{s^\prime}^\dagger(q_0,\vec{q})+e^{aq_0\frac{\partial}{\partial p_i}p_i-ap_0\frac{\partial}{\partial q_i}q_i}b_{s^\prime}^\dagger(q_0,\vec{q})b_s^\dagger(p_0,\vec{p})=0,
\ee
\be
d_s(p_0,\vec{p})d_{s^\prime}(q_0,\vec{q})+e^{-aq_0\frac{\partial}{\partial p_i}p_i+ap_0\frac{\partial}{\partial q_i}q_i}d_{s^\prime}(q_0,\vec{q})d_s(p_0,\vec{p})=0.\label{57}
\ee
From this, we can easily derive the following relations
\bea
b_s(p_0,e^{-\frac{aq_0}{2}}\vec{p})b_{s^\prime}^\dagger(q_0,e^{-\frac{ap_0}{2}}\vec{q})+e^{3a(q_0+p_0)}b_{s^\prime}^\dagger(q_0,e^{\frac{ap_0}{2}}\vec{q})b_s(p_0,e^{\frac{aq_0}{2}}\vec{p})\nonumber\\=\delta^3(p-q)\delta_{ss^\prime},
\eea
\bea
d_s^\dagger(p_0,e^{\frac{aq_0}{2}}\vec{p})d_{s^\prime}(q_0,e^{\frac{ap_0}{2}}\vec{q})+e^{-3a(q_0+p_0)}d_{s^\prime}(q_0,e^{-\frac{ap_0}{2}}\vec{q})d_s^\dagger(p_0,e^{-\frac{aq_0}{2}}\vec{p})\nonumber\\=\delta^3(p-q)\delta_{ss^\prime},
\eea
\be
b_s(p_0,e^{\frac{aq_0}{2}}\vec{p})b_{s^\prime}(q_0,e^{-\frac{ap_0}{2}}\vec{q})+e^{3a(-q_0+p_0)}b_{s^\prime}(q_0,e^{\frac{ap_0}{2}}\vec{q})b_s(p_0,e^{-\frac{aq_0}{2}}\vec{p})=0,
\ee
\be
d_s^\dagger(p_0,e^{-\frac{aq_0}{2}}\vec{p})d_{s^\prime}^\dagger(q_0,e^{\frac{ap_0}{2}}\vec{q})+e^{3a(q_0-p_0)}d_{s^\prime}^\dagger(q_0,e^{-\frac{ap_0}{2}}\vec{q})d_s^\dagger(p_0,e^{\frac{aq_0}{2}}\vec{p})=0,
\ee
\be
b_s^\dagger(p_0,e^{-\frac{aq_0}{2}}\vec{p})b_{s^\prime}^\dagger(q_0,e^{\frac{ap_0}{2}}\vec{q})+e^{3a(q_0-p_0)}b_{s^\prime}^\dagger(q_0,e^{-\frac{ap_0}{2}}\vec{q})b_s^\dagger(p_0,e^{\frac{aq_0}{2}}\vec{p})=0,
\ee
\be
d_s(p_0,e^{\frac{aq_0}{2}}\vec{p})d_{s^\prime}(q_0,e^{-\frac{ap_0}{2}}\vec{q})+e^{3a(-q_0+p_0)}d_{s^\prime}(q_0,e^{\frac{ap_0}{2}}\vec{q})d_s(p_0,e^{-\frac{aq_0}{2}}\vec{p})=0.
\ee
As in\cite{12}, we define a new product between the creation and annihilation operators as
\be
b_s(p_0,\vec{p})~o~b_{s^\prime}^\dagger(q_0,\vec{q})=e^{-\frac{3a}{2}(q_0+p_0)}b_s(p_0,e^{-\frac{aq_0}{2}}\vec{p})b_{s^\prime}^\dagger(q_0,e^{-\frac{ap_0}{2}}\vec{q})
\ee
\be
d_s^\dagger(p_0,\vec{p})~o~d_{s^\prime}(q_0,\vec{q})=e^{\frac{3a}{2}(q_0+p_0)}d_s^\dagger(p_0,e^{\frac{aq_0}{2}}\vec{p})d_{s^\prime}(q_0,e^{\frac{ap_0}{2}}\vec{q})
\ee
\be
b_s(p_0,\vec{p})~o~b_{s^\prime}(q_0,\vec{q})=e^{-\frac{3a}{2}(-q_0+p_0)}b_s(p_0,e^{\frac{aq_0}{2}}\vec{p})b_{s^\prime}(q_0,e^{-\frac{ap_0}{2}}\vec{q})
\ee
\be
d_s^\dagger(p_0,\vec{p})~o~d_{s^\prime}^\dagger(q_0,\vec{q})=e^{\frac{3a}{2}(-q_0+p_0)}d_s^\dagger(p_0,e^{-\frac{aq_0}{2}}\vec{p})d_{s^\prime}^\dagger(q_0,e^{\frac{ap_0}{2}}\vec{q})
\ee
\be
b_s^\dagger(p_0,\vec{p})~o~b_{s^\prime}^\dagger(q_0,\vec{q})=e^{\frac{3a}{2}(-q_0+p_0)}b_s^\dagger(p_0,e^{-\frac{aq_0}{2}}\vec{p})b_{s^\prime}^\dagger(q_0,e^{\frac{ap_0}{2}}\vec{q})
\ee
\be 
d_s(p_0,\vec{p})~o~d_{s^\prime}(q_0,\vec{q})=e^{-\frac{3a}{2}(-q_0+p_0)}d_s(p_0,e^{\frac{aq_0}{2}}\vec{p})d_{s^\prime}(q_0,e^{-\frac{ap_0}{2}}\vec{q}).
\ee
Using this new $o$-product rule, we can re-write Eqns.(\ref{52}-\ref{57}) in a compact form as
\be
\lbrace b_s(p_0,\vec{p})~,~b_{s^\prime}^\dagger(q_0,\vec{q})\rbrace_o=\delta^3(p-q)\delta_{ss^\prime},\label{70}
\ee
\be
\lbrace d_s(p_0,\vec{p})~,~d_{s^\prime}^\dagger(q_0,\vec{q})\rbrace_o=\delta^3(p-q)\delta_{ss^\prime},
\ee
\be 
\lbrace d_s(p_0,\vec{p})~,~d_{s^\prime}(q_0,\vec{q})\rbrace_o=\lbrace d_s^\dagger(p_0,\vec{p})~,~d_{s^\prime}^\dagger(q_0,\vec{q})\rbrace_o=0
\ee
\be 
\lbrace b_s(p_0,\vec{p})~,~b_{s^\prime}(q_0,\vec{q})\rbrace_o=\lbrace b_s^\dagger(p_0,\vec{p})~,~b_{s^\prime}^\dagger(q_0,\vec{q}\vec{q})\rbrace_o=0 .\label{73}
\ee
These are the deformed anticommutation relations for the fermionic creation and annihilation operators and they have the same form as in the commutative case. Note that in limit $a\rightarrow 0$, we get back the commutative result. These creation and annihilation operators satisfying the o-product rule are the ones appearing in the mode decomposition of fermionic field (see Eqns. (\ref{50},\ref{51})) satisfying the $\kappa$-deformed Dirac Eqn.(\ref{24}). 

\section{Conclusion}

In this paper, we started with the second quantized Dirac field in the $\kappa$-Minkowski space-time and derived the deformed algebra of creation and annihilation operators corresponding to the fermions described by this Dirac field. This is done by demanding the compatibility of the flip operation and the action of the Hopf algebra, which is the symmetry algebra corresponding to the $\kappa$-deformed Minkowski space-time. The Hopf algebra structure do modify the notion of the flip operator and this changes the notion of statistics. Further by using a modified product, we show that the deformed algebra can be cast in the conventional form. In the limit $a\rightarrow 0$, we recover the well known commutative result. We have seen here the Fermi-Dirac statistics is deformed in the $\kappa$-space-time. This will have the implications in the particle physics. We are studying effects of the deformed statistics on Unruh effect involving fermions in $\kappa$-space-time\cite{eh1,eh}. Work along these lines in 
progress and reported separately.
\begin{flushleft}
\begin{large}
\textbf{Acknowledgments}
\end{large}
\end{flushleft}
Author thank E. Harikumar for suggestions and discussions. He is also thankful to UGC(India) for financial support through Rajiv Gandhi National Fellowship(F1-17.1/2011-12/RGNF-SC-UTT-4237).

\end{document}